\documentstyle [fleqn,12pt] {article}
\def\mscript#1{\mbox{\scriptsize$#1$}}

\begin{document}
\pagestyle{empty}
\begin{flushleft}
PACS No. 11.10G;  11.17;  12.10
\end{flushleft}
\begin{flushright}
RUP-96-4 \\
July, 1996 \\
\end{flushright}
\vspace{1cm}
\begin{center}
Thermofield Dynamics of the Heterotic String\\
 --- Global Phase Structure --- \\
\vspace{1cm} 
H. Fujisaki,\footnote[1] {e-mail: sano@rikkyo.ac.jp} 
K.Nakagawa\footnote[7] {e-mail: nakagawa@hoshi.ac.jp} and S. Sano$^*$

\vspace{5mm}

$^*$ Department of Physics, Rikkyo University, Tokyo 171\\
$^{**}$ Faculty of Pharmaceutical Sciences, Hoshi University, Tokyo 142\\
\vspace*{3cm}
{\bf ABSTRACT}
\end{center}

\indent Physical aspects of the thermofield dynamics of the $D = 10$ 
heterotic thermal string theory are exemplified through the infrared 
behaviour of the one-loop dual symmetric cosmological constant in 
association with the global phase structure of the thermal string 
ensemble.  
\newpage
\pagestyle{plain}
\setcounter{page}{1}
\indent Elaboration of thermal string theories based upon the thermofield 
dynamics (TFD) \cite{umezawa} has gradually turned out to be a good 
practical subject of high energy physics \cite{leblanc1} - 
\cite{nakagawa}.  In a previous paper of ourselves \cite{fujisaki4}, we 
have succeeded in shedding some light upon the global phase structure of 
the thermal string excitation in proper reference to the thermal duality 
relation \cite{obrien}, \cite{atick} for the $D = 26$ closed bosonic 
thermal string theory within the TFD framework.  In the present 
communication, physical aspects of the $D = 10$ heterotic thermal string 
theory based upon the TFD algorithm are exemplified through the infrared 
behaviour of the one-loop cosmological constant in respect of the thermal 
duality symmetry.  The global phase structure of the TFD thermal string 
amplitude is then examined $\grave{a}\: la$ O'Brien and Tan 
\cite{obrien} on the basis of the thermal stability of modular invariance.

Let us start with describing the one-loop cosmological constant 
$\Lambda(\beta)$ at any finite temperature $\beta^{-1} = kT$ as
\vspace{5mm}
\begin{equation}
\Lambda (\beta) = \frac{\alpha^\prime}{2} \lim_{\mu^2 \rightarrow 0} {\rm 
Tr} 
\left[ \int_{\infty}^{\mu^2} dm^2 \left( \Delta^\beta_B (p, P; m^2) + 
\Delta^{\beta}_{F} (p, P; m^2) \right) \right]
\end{equation}

\vspace{5mm}
\noindent for the $D = 10$ heterotic thermal string theory in the TFD 
framework, 
where the string tension $\sigma$ is expressed in terms of the slope 
parameter $\alpha^\prime$ as $\sigma = 1/2\pi \alpha^\prime$, $p^\mu$ 
reads loop momentum and $P^I$ lie on the even self-dual root lattice $L = 
\Gamma_8 \times \Gamma_8$ for the exceptional group $G = E_8 \times 
E_8$ \cite{green}, respectively.  Here the thermal propagator 
$\Delta^\beta_{B[F]} (p, P; m^2)$ of the free closed bosonic [fermionic] 
string is expressed at $D = 10$ as
 
\begin{eqnarray}
\lefteqn{\Delta^\beta_{B[F]}(p, P; m^2) = \int_{-\pi}^{\pi} 
\frac{d\phi}{4\pi} \; {\rm e}^{i\phi \left( N - \alpha - \bar{N} + 
\bar{\alpha} -1/2 \cdot \sum_{I=1}^{16} (P^I)^2 \right) }}
\nonumber \\
& & \times \Biggl( \left[ \raisebox{-1ex}{$\stackrel {\textstyle 
+}{\mscript{[}-\mscript{]}}$} \int_{0}^{1} dx + \frac{1}{2} \sum_{n=0}^{\infty} \frac{\delta [\alpha^\prime 
/2 \cdot p^2 + \alpha^\prime /2 \cdot m^2 + 2(n - \alpha)]}{{\rm e}^{\beta 
|p_0|} \raisebox{-1ex}{$\stackrel{\textstyle -}{\mscript{[}+\mscript{]}}$} \; 1} 
\oint_{c} dx \right] \nonumber \\
& & \times x^{\alpha^\prime /2 \cdot p^2 + N - \alpha + \bar{N} - 
\bar{\alpha} + 1/2 \cdot \sum_{I=1}^{16} (P^I)^2 + \alpha^\prime /2 \cdot 
m^2 - 1} \Biggr) \quad ,
\end{eqnarray}

\vspace{5mm}
\noindent where $N$ [$\bar{N}$] denotes the number operator of the right- 
[left-] 
moving mode, the intercept parameter $\alpha$ [$\bar{\alpha}$] of the 
right- [left-] mover is eventually fixed at $\alpha = 0$  
$[\bar{\alpha} = 1]$ and the contour $c$ is taken as the unit circle 
around the origin.  The $D = 10$ thermal cosmological constant $\Lambda 
(\beta)$ is immediately reduced to
\vspace{5mm}
\begin{eqnarray}
\Lambda (\beta) & = & - \frac{i\alpha^\prime}{4} \lim_{\mu^2 
\rightarrow 0} \int_{\infty}^{\mu^2} dm^2 \int_{-\infty}^{\infty} 
\frac{d^Dp}{(2\pi)^D} \sum_{n=0}^\infty \delta \left[ \frac{\alpha^\prime}
{2} p^2 + \frac{\alpha^\prime}{2} m^2 + 2(n - \alpha) \right] \nonumber \\
& & \times \left( \frac{1}{{\rm e}^{\beta|p_0|} - 1} + \frac{1}{{\rm 
e}^{\beta|p_0|} + 1} \right) \sum_{P^I \in L} \int_{-\pi}^{\pi} 
\frac{d\phi}{4\pi} {\rm e}^{-i\phi \left( \alpha -\bar{\alpha}+1/2\cdot 
\sum_{I=1}^{16} (P^I)^2 \right)} \nonumber \\
& & \times \oint_c dx x^{\alpha^\prime /2 \cdot p^2 - \alpha - 
\bar{\alpha} + 1/2 \cdot \sum_{I=1}^{16} (P^I)^2 + \alpha^\prime /2 
\cdot m^2-1} \nonumber \\
& & \times {\rm tr} \left[ {\rm e}^{i\phi(N-\bar{N})} x^{N+\bar{N}} 
\right] \quad.  
\end{eqnarray}

\vspace{5mm}
Let us turn our attention to explicit calculation of the $D = 10$ thermal 
amplitude $\Lambda (\beta)$.  We are then eventually led to 
the modular parameter integral representation of $\Lambda (\beta)$ at $D = 10$ 
as follows \cite{obrien}:
\vspace{5mm}
\begin{equation}
\Lambda (\beta) = -8(2\pi \alpha^\prime)^{-D/2} \int_{E} 
\frac{d^2\tau}{2\pi \tau_2^2} \; K_h (\bar{\tau}, \tau; D) \sum_{\ell \in Z; {\rm odd}} \exp \left[ 
- \frac{\beta^2}{4\pi \alpha^\prime \tau_2}\; \ell^2 \right] \quad, 
\label{eq:Lambda2}
\end{equation}
where
\begin{eqnarray}
K_h(\bar{\tau}, \tau; D) & = & (2\pi \tau_2)^{-(D-2)/2}\; {\rm 
e}^{2\pi i \bar{\tau}} \left[ 1 + 480 \sum_{m=1}^{\infty} 
\sigma_7(m)\bar{z}^m \right] \nonumber \\
& & \times \prod_{n=1}^{\infty} (1 - \bar{z}^n)^{-D-14} \left( 
\frac{1 + z^n}{1 - z^n} \right)^{D-2} \quad ,
\label{eq:K} 
\end{eqnarray}

\vspace{5mm}
\noindent $\stackrel{[\normalsize{-}]}{\tau} = \tau_1 
\raisebox{-1ex}{$\stackrel{\normalsize{+}}{\mscript{[}\!-\!\mscript{]}}$} 
i\tau_2$, 
$z = x {\rm e}^{i\phi} = {\rm 
e}^{2\pi i \tau}$, $\bar{z} = x{\rm e}^{-i \phi} = {\rm e}^{-2\pi i 
\bar{\tau}}$, $\alpha \; [\bar{\alpha}]$ has been fixed at $\alpha = 
0 \; [\bar{\alpha} = 1]$, $E$ means the half-strip integration 
region in the complex $\tau$ plane, i.e. $-1/2 \leq \tau_1 \leq 1/2$; 
$\tau_2 > 0$, and the full use has been made of an explicit expression of 
the theta 
function $\Theta_{\Gamma_8 \times \Gamma_8}$ of the root lattice 
$\Gamma_8 \times \Gamma_8$ \cite{green}.  Accordingly, the $D = 10$ 
thermal amplitude $\beta \Lambda (\beta)$ is identical in every detail 
with the ``$E$-type'' representation of the thermo-partition function 
$\Omega_h (\beta)$ of the heterotic string in ref. \cite{obrien} as 
required from the equivalence of the thermal cosmological constant and 
the free energy amplitude.  As can readily be envisaged from eqs. 
($\!$~\ref{eq:Lambda2}) and ($\!$~\ref{eq:K}), the ``$E$-type'' thermal amplitude $\Lambda 
(\beta)$ 
is not modular invariant and consequently is annoyed with ultraviolet 
divergences at the endpoint $\tau_2 \sim 0$ for any value of $\beta$.  

Let us now postulate the one-loop dual symmetric thermal cosmological 
constant $\bar{\Lambda} (\beta; D)$ at any space-time dimension $D$  as an integral over the 
fundamental domain $F$, i.e. $-1/2 
\leq \tau_1 \leq 1/2 \: ; \tau_2 > 0 \: ; |\tau| > 1$, of the modular 
group 
$SL(2, Z)$ as follows \cite{obrien} :  
\begin{equation}
\bar{\Lambda} (\beta; D) = - 
\frac{16}{\beta} (2\pi \alpha^\prime)^{-D/2} \sum_{(\sigma, \rho)} 
\int_{F} 
\frac{d^2\tau}{2\pi \tau_2^2}\; B(\bar{\tau}, \tau; 
D) A_{\sigma \rho} (\tau; D) D_{\sigma \rho} (\bar{\tau}, \tau; \beta) \quad,
\end{equation}
where
\vspace{5mm}
\begin{eqnarray}
B(\bar{\tau}, \tau; D) & = & - \frac{1}{8} (2 \pi \tau_2)^{-(D-2)/2}\;  
\bar{z}^{-(D+14)/24} z^{-(D-2)/24} \nonumber \\
& & \times \left[ 1 + 480 \sum_{m=1}^{\infty} \sigma_7 (m) 
\bar{z}^m \right] \prod_{n=1}^{\infty} (1 - \bar{z}^n)^{-D-14} 
(1 - z^n)^{-D+2}, 
\end{eqnarray}

\begin{eqnarray}
\left( \begin{array}{c}
A_{+-}(\tau; D) \rule[-2mm]{0mm}{8mm} \\ A_{-+}(\tau; D) \rule[-2mm]{0mm}{8mm} \\ A_{--}(\tau; D) 
\end{array} \right) 
= 8 \left( \frac{\pi}{4} \right) ^{(D-2)/6} 
\left( \begin{array}{l}
-[\theta_2(0, \tau)/\theta_{1}^{\prime}(0, \tau)^{1/3}]^{(D-2)/2}
\rule[-2mm]{0mm}{8mm} \\
-[\theta_4(0, \tau)/\theta_{1}^{\prime}(0, \tau)^{1/3}]^{(D-2)/2}
\rule[-2mm]{0mm}{8mm} \\
\; [\theta_3(0, \tau)/\theta_{1}^{\prime}(0, \tau)^{1/3}]^{(D-2)/2}
\rule[-2mm]{0mm}{8mm}
\end{array} \right) \quad ,
\end{eqnarray}

\vspace{5mm}
\begin{equation}
D_{\sigma \rho}(\bar{\tau}, \tau; \beta) = 
C_\sigma^{(+)}(\bar{\tau}, \tau; \beta) 
+ \rho C_\sigma^{(-)}(\bar{\tau}, \tau; \beta) \quad,
\end{equation}

\vspace{5mm}
\begin{equation}
C_\sigma^{(\gamma)}(\bar{\tau}, \tau; \beta) = (4\pi^2\alpha^\prime 
\tau_2)^{1/2} \sum_{(p, q)} \exp \left[ - \frac{\pi}{2} \left( 
\frac{\beta^2}{2\pi^2\alpha^\prime} p^2 + \frac{2\pi^2\alpha^\prime}
{\beta^2} q^2 \right) \tau_2 + i\pi pq \tau_1 \right] ,
\end{equation}

\vspace{5mm}
\noindent the signatures $\sigma, \rho$ and $\gamma$ read $\sigma, \rho = +, 
- ; \; -, + ; \; -, -$ and $\gamma = +, -$, respectively, the summation 
over $p \; [q]$ is restricted by $(-1)^p = \sigma \; [(-1)^q = \gamma]$ 
and the 
explicit use has been made of the Jacobi theta functions $\theta_j(0, 
\tau);\; j = 1, 2, 3, 4$ as well as the Poisson resummation formula.  It 
is 
almost needless to mention that the $D = 10$ thermal amplitude 
$\beta\bar{\Lambda}(\beta; D = 10)$ is literally reduced to the 
``$D$-type'' representation of the thermo-partition function 
$\Omega_h(\beta)$ in ref. \cite{obrien} which in turn guarantees 
$\bar{\Lambda}(\beta; D = 10) = \Lambda (\beta)$ as expected from 
self-consistency.  Let us now examine 
the algebraic structure of the ``$D$-type'' thermal amplitude 
$\bar{\Lambda}(\beta; D)$ with the arithmetic aid of Appendix B in 
ref. \cite{obrien}.  Typical theoretical observations are as follows:  The 
scalar product $\sum_{(\sigma, \rho)} A_{\sigma \rho} D_{\sigma \rho}$ is 
invariant under permutations of the signature, irrespective of the values 
of $\beta$ and $D$, not only for the shifting transformation $\tau 
\rightarrow \tau + 1$ but also for the inversion $\tau \rightarrow 
-\tau^{-1}$.  In addition, $B(\bar{\tau}, \tau; D)$ is invariant, 
irrespective of the value of $D$, under the action of any modular 
transformation.  Accordingly, the ``$D$-type'' thermal amplitude 
$\bar{\Lambda}(\beta; D)$ is manifestly modular invariant and thus 
free of ultraviolet divergences for any value of $\beta$ and $D$.  As a 
matter of fact, moreover, the generalized duality symmetry \cite{obrien} 
$C_\sigma^{(\gamma)}(\bar{\tau}, \tau; \beta) = 
C_\gamma^{(\sigma)}(\bar{\tau}, \tau; \tilde{\beta})$ holds for any value 
of $D$, where $\tilde{\beta} = 2\pi^2 \alpha^\prime /\beta$.  If and only 
if $D = 10$, on the other hand, the scalar 
product $\sum_{(\sigma, \rho)} A_{\sigma \rho} D_{\sigma \rho}$ is 
invariant under the thermal duality transformation $\beta \leftrightarrow 
\tilde{\beta}$ as a simple and natural consequence of the Jacobi identity 
$\theta_2^4 - \theta_3^4 + \theta_4^4 = 0$ for the theta functions.  We 
are then led to conclude that the thermal duality relation $\beta 
\bar{\Lambda}(\beta; D) = 
\tilde{\beta}\bar{\Lambda}(\tilde{\beta}; D)$ is manifestly broken 
for the ``$D$-type'' thermal amplitude $\bar{\Lambda}(\beta; D \neq 
10)$ off the critical dimension.  

Let us recall to our remembrance that $\theta_{1}^{\prime -1/3} \sim {\rm 
e}^{\pi \tau_2/12};\; \theta_2 \sim 0;\; \theta_3 \sim 1;\; \theta_4 \sim 
1$ 
near $\tau_2 \rightarrow \infty$.  The infrared behaviour of the 
``$D$-type'' thermal cosmological constant $\bar{\Lambda}(\beta; D)$ 
is then asymptotically described at $\tau_2 \rightarrow \infty$ as 
\cite{obrien} 
\begin{equation}
\bar{\Lambda}(\beta; D) = - \frac{16}{\beta} \; (2\pi \alpha^\prime)^{-D/2} 
\int_{F} \frac{d^2\tau}{2\pi \tau_2^2} \; B(\bar{\tau}, \tau; D) 
\left[ A_{-+}(\tau; D) - A_{--}(\tau; D) \right] C_-^{(-)}(\bar{\tau},
 \tau; \beta)
\end{equation}

\vspace{5mm}
\noindent which is in turn paraphrased into the form 
\begin{eqnarray}
\bar{\Lambda}(\beta; D) & = & - 64 \sqrt{2}\: (8\pi^2 \alpha^\prime)^{-D/2} 
\sum_{(p, q)} \int_{- \frac{1}{2}}^{\frac{1}{2}} d\tau_1 \exp [i\pi 
pq\tau_1] \, \sqrt{\frac{\tilde{\beta}}{\beta}} 
\int_{\sqrt{1-\tau_1^2}}^{\infty} d\tau_2 \; \tau_2^{-(D+1)/2} \nonumber \\
& & \times \exp \left[ - \frac{\pi}{2} \; \tau_2 \left( 
\frac{\beta}{\tilde{\beta}} \; p^2 + \frac{\tilde{\beta}}{\beta} \; q^2 - 
\frac{5}{12} (D - 10) - 6 \right) \right] \quad ,
\label{eq:Lambda3}
\end{eqnarray}

\vspace{5mm}
\noindent where $p, q = \pm 1; \pm 3; \pm 5; \cdots$.  As can easily be seen 
from 
eq. ($\!$~\ref{eq:Lambda3}), uniform convergence of the ``$D$-type'' thermal
 amplitude $\bar{\Lambda}(\beta; D)$ is assured at any value of $\beta$ if and 
only if $D < 2/5$.  Of principal concern with us is the case $D = 10$, 
anyhow.  Infrared convergence of the $D = 10$ ``$D$-type'' TFD amplitude 
$\bar{\Lambda}(\beta; D = 10)$ is then guaranteed if and only if 
either $(2 + \sqrt{2})\pi \sqrt{\alpha^\prime} = \beta_H < \beta < 
\infty$ or $0 < \beta < \tilde{\beta}_H = (2 - \sqrt{2}) \pi 
\sqrt{\alpha^{\prime}}$, where $\beta_H$ 
$[\tilde{\beta}_H]$ reads the inverse [dual] Hagedorn temperature of the 
heterotic thermal string.  Explicit calculation of the $\tau_2$ integral 
in eq. ($\!$~\ref{eq:Lambda3}) is readily performed for the case $D < 
2/5$ and yields
\begin{eqnarray}
\bar{\Lambda}(\beta; D) & = & - \frac{128}{\sqrt{\pi}} (16\pi 
\alpha^{\prime})^{-D/2} \sum_{(p,q)} \int_{-\frac{1}{2}}^{\frac{1}{2}} 
d\tau_1 \exp [i\pi pq\tau_1] \nonumber \\
& & \times \sqrt {\frac{\tilde{\beta}}{\beta}} \left( \frac{\beta}{\tilde{\beta}} \; 
p^2 + \frac{\tilde{\beta}}{\beta} \; q^2 - \frac{5}{12} (D - 10) - 6 
\right)^{(D-1)/2} \nonumber \\
& & \times \Gamma \left[ - \frac{D - 1}{2} , \frac{\pi}{2} 
\sqrt{1 - \tau_{1}^{2}} \left( \frac{\beta}{\tilde{\beta}} \; p^2 + 
\frac{\tilde{\beta}}{\beta} \; q^2 - \frac{5}{12} (D - 10) - 6 \right) 
\right] \quad ,
\label{eq:Lambda4}
\end{eqnarray}

\vspace{5mm}
\noindent irrespective of the value of $\beta$, where $\Gamma$ is the 
incomplete gamma function of the second kind.  We are now in the position 
to carry out the dimensional regularization in the sense of analytic 
continuation of the TFD amplitude ($\!$~\ref{eq:Lambda4}).  The 
right-hand side of eq. ($\!$~\ref{eq:Lambda4}) indeed obeys the thermal 
duality symmetry $\beta \bar{\Lambda}(\beta; D) = 
\tilde{\beta}\bar{\Lambda}(\tilde{\beta}; D)$ and brings forth the 
correct analytic continuation from $D < 2/5$ to higher values of $D$, 
i.e. $D = 10$.  We can therefore define the dimensionally regularized, 
$D = 10$ one-loop dual symmetric thermal cosmological constant 
$\hat{\Lambda}(\beta)$ by 
\begin{eqnarray}
\hat{\Lambda}(\beta) & = &  -\frac{2}{\beta} (8\pi 
\alpha^\prime)^{-(D-1)/2} \sum_{(p, q)} 
\int_{- \frac{1}{2}}^{\frac{1}{2}} d\tau_1 \exp[i\pi pq \tau_1] \nonumber 
\\
& & \times \left( \frac{\beta^2}{2\pi^2\alpha^\prime} \: p^2 + 
\frac{2\pi^2\alpha^\prime}
{\beta^2} \: q^2 - 6 \right) ^{(D-1)/2} \nonumber \\
& & \times \Gamma \left[ - \frac{D - 1}{2}\: ,\; \frac{\pi}{2} \sqrt{1 - 
\tau_1^2} \left( \frac{\beta^2}{2\pi^2 \alpha^\prime} \: p^2 + 
\frac{2\pi^2 
\alpha^\prime}{\beta^2} \: q^2 - 6 \right) \right]\; ;\quad D = 10
\label{eq:Lambda5}
\end{eqnarray}

\vspace{5mm}
\noindent which manifestly satisfies the thermal duality relation $\beta 
\hat{\Lambda}(\beta) = \tilde{\beta} \hat{\Lambda}(\tilde{\beta})$.  It 
must be emphasized that the present dimensional regularization based upon 
the TFD algorithm is in full accordance with the thermal stability of the 
fundamental properties such as modular invariance.  

Let us next examine the singularity structure of the dimensionally 
regularized, $D = 10$ dual symmetric thermal amplitude 
$\hat{\Lambda}(\beta)$.  The position of the singularity $\beta_{|p|, 
|q|}$ is determined by solving $\beta/\tilde{\beta} \cdot p^2 + 
\tilde{\beta}/\beta \cdot q^2 - 6 = 0$ for every allowed $(p, q)$ in eq. 
($\!$~\ref{eq:Lambda5}).  We then obtain two sets of solutions 
with $|pq| \leq 3$ as follows: (i) $\beta_{1, 1} = \beta_H = (\sqrt{2} + 
1) \pi \sqrt{2\alpha^\prime}\: ; \; \tilde{\beta}_{1, 1} = \tilde{\beta}_H 
= 
(\sqrt{2} - 1)\pi \sqrt{2\alpha^\prime}\:$ , (ii) $\beta_{1, 3} = 
\tilde{\beta}_{3, 1} = \sqrt{3} \pi \sqrt{2\alpha^\prime}\: ; \; \beta_{3, 
1} 
= \tilde{\beta}_{1, 3} = 1/\sqrt{3} \cdot \pi 
\sqrt{2\alpha^\prime}\:$ .  In 
particular, $\beta_{1, 1}$ and $\tilde{\beta}_{1, 1}$ form the leading 
branch points of the square root type at $\beta_H$ and $\tilde{\beta}_H$, 
respectively.  Moverover, $\beta_H^{-1}$ $[\tilde{\beta}_H^{-1}]$ 
represents the lowest temperature singularity for the physical $\beta$ 
[dual $\tilde{\beta}$] channel.  Both $\beta_{1, 3}$ and $\beta_{3, 1}$ 
are ordinary points and consequently left out of consideration.  It is of 
practical importance to note that there exists no self-dual leading 
branch point at $\beta_0 = \tilde{\beta}_0 = \pi \sqrt{2\alpha^\prime}$ 
as well as any non-leading branch point on the physical sheet of 
the inverse temperature.  The present theoretical observation based upon 
the TFD free energy amplitude of the $D = 10$ heterotic thermal string 
yields a striking contrast to the previous argument by ourselves 
\cite{fujisaki4} for the $D = 26$ closed bosonic thermal string theory in 
the TFD framework.  We are now in the position to touch upon the global 
phase structure of the $D = 10$ heterotic thermal string ensemble.  
Analysis is performed $\grave{a}\: la$ ref. \cite{obrien}, ref. 
\cite{leblanc3} and ref. \cite{deo} through the microcanonical ensemble 
paradigm outside the analyticity domain of the canonical ensemble.  There 
will then appear three phases in the sense of the thermal duality 
symmetry as follows \cite{fujisaki4}, \cite{obrien}, \cite{leblanc3}: (i) 
the $\beta$ channel canonical phase in the domain $(2 + \sqrt{2})\pi \sqrt
{\alpha^\prime} = \beta_H \leq \beta < \infty$, (ii) the dual 
$\tilde{\beta}$ channel canonical phase in the domain $0 < \beta \leq 
\tilde{\beta}_H = (2 - \sqrt{2})\pi \sqrt{\alpha^\prime}$ and 
(iii) the self-dual microcanonical phase in the domain $\tilde{\beta}_H < 
\beta 
< \beta_H$.  In sharp contrast to the global phase structure of the $D = 
26$ closed bosonic thermal string ensemble \cite{fujisaki4}, however, 
there 
will occur no effective splitting of the microcanonical region because of 
the absence of the self-dual branch point at $\beta_0 = \tilde{\beta}_0 
= \pi \sqrt{2\alpha^\prime}$ as well as any secondary singularity.  As a 
consequence of the self-duality of the microcanonical phase, therefore, 
it may be possible to claim that the so-called maximum temperature of 
the $D = 10$ heterotic string excitation is asymptotically described at 
least at the one-loop level as $\tilde{\beta}_H^{-1}$ $[\beta_H^{-1}]$ in 
replacement of $\beta_0^{-1} = \tilde{\beta}_0^{-1}$ for the physical 
$\beta$ [dual $\tilde{\beta}$] channel.  It seems almost needless to 
mention 
that the fruitful thermodynamical investigation of string excitations 
will be prerequisite for the real solid substantiation of the present 
novel hypothesis on the ``true'' maximum temperature in proper reference 
to the global phase structure of the $D = 10$ heterotic thermal string 
ensemble.  It is hoped that we can shed some light upon this unfathomable 
subject with the new-fashioned aid of the $D$-brane paradigm \cite{mozo} 
in a future communication.  

\end{document}